\documentclass[twocolumn]{article}
\usepackage{geometry}
\usepackage{caption}
\usepackage{graphicx}
\usepackage{amsmath}
\usepackage{amssymb}
\usepackage{textcomp}
\usepackage[dvipsnames]{xcolor}
\usepackage{authblk}
\usepackage{datetime}
\usepackage{gensymb}
\usepackage{wrapfig}
\usepackage{booktabs}
\usepackage{ulem}
\usepackage[numbers]{natbib}
\usepackage{hyperref}
\hypersetup{
    citecolor = blue,
    filecolor = black,
    urlcolor = blue,
    colorlinks = true, 
    linktoc = all,     
    linkcolor = blue,  
}

\title{High-resistivity niobium nitride films for unity-efficiency SMSPDs at telecom wavelengths and beyond}
\author[1]{\large{Philipp Zolotov$^{*1}$, Sergey Svyatodukh$^{1,2}$, Alexander Divochiy$^3$, Vitaliy Seleznev$^{2,3}$,\newline and Gregory Goltsman$^{1,2,3}$}
\\
\vspace{0.5em}
\textit{\large{$^1$Moscow Institute of Electronics and Mathematics}}
\\
\textit{\large{Tallinskaya Ulitsa 34, 123592, Moscow, Russia}}
\vspace{0.5em}
\vspace{0.5em}\\
\textit{\large{$^2$Moscow Pedagogical State University}}\\
\textit{\large{Malaya Pirogovskaya Ulitsa 1/1, 119435, Moscow, Russia}}\\
\vspace{0.5em}
\vspace{0.5em}
\textit{\large{$^3$Superconducting Nanotechnology (SCONTEL)}}\\
\textit{\large{Derbenevskaya Naberezhnaya 11kA, 115114, Moscow, Russia}}\\
\vspace{0.5em}
\small{$^*$pizolotov@ya.ru}
}
\date{\today}

\begin{document}

\twocolumn[
\begin{@twocolumnfalse}
\maketitle
\begin{abstract}
The sensitive element of superconducting single-photon detectors made in the form of a microstrip promise to resolve significant limitations caused by their typical design. However, attention should be paid to the problem of deterioration of the detection efficiency of devices with an increase of the width of the superconducting strip from nano- to microscale. This article demonstrates a possibility of achieving highly saturated detection efficiency of superconducting microstrip single-photon detectors by using high-resistivity niobium nitride films. The approach opens the way for employing fundamentally improved experimental devices.

\vspace{3em}
\end{abstract}
\end{@twocolumnfalse}
]

\setcounter{tocdepth}{1}
\setcounter{secnumdepth}{4}

Superconducting single-photon detectors (SSPDs) gained a significant attention soon after pioneering work back in 2001\cite{gol2001picosecond}. Nowadays these detectors are known for their wide operation range from ultra-violet to mid-infrared, high counting rates of the order of 1 GHz, 3 ps jitter, and low dark count rates as low as a count per day \cite{wollman2017uv,marsili2012efficient,vetter2016cavity,korzh2020demonstration,goltsman2004nano}. However, in practical applications in fields of quantum optics, astronomy, and others, preference is usually given to another significant metric -- photon detection efficiency (DE)\cite{moreau2019imaging,unternahrer2018super,ozana2021superconducting,wang2022twin}. Practical realizations of SSPDs require a specific design of the sensitive element of the detectors that is fabricated as a narrow and long meandering nano-sized strip or \textit{nanowire}. Recently, Colangelo \textit{et al.} demonstrated that superconducting \textit{nanowire} single-photon detectors (SNSPDs) reach unity DE not only in near-, but also in middle-infrared range\cite{colangelo2022large}. This achievement was granted by nanowires as narrow as 50 nm with a total length of the order of several hundred microns that were fabricated from tungsten silicide film. However, such a peculiar design significantly limits detector's dead time and temporal resolution due to high kinetic inductance\cite{kerman2006kinetic}. Moreover, it restricts device fabrication capacities and raises the bar for technological requirements.

The essence of SSPD's photoresponse is the transition of the strip cross section from superconducting to normal state. Emerging after photon absorption hotspot, which initiates this transition, has a size of about several tens of nanometers, which explains the desire to maximize its influence by making narrow strips. However, upon closer examination, it turns out that a vortex -- antivortex pair (or a single vortex that appears when a spot occurs at the edge of the strip) generated in the hotspot can lead to a formation of normal cross section almost without any reference to strip width. This conclusion was drawn by Zotova and Vodolazov in 2012 and launched a new branch of research involving micrometer-scale strips\cite{zotova2012photon,vodolazov2017single}. Later, in 2018 a fundamentally new experimental results bode a solution to described downsides of the mainstream SNSPD design\cite{korneeva2018optical}. Authors demonstrated that short structures as wide as 5 $\mu$m fabricated from niobium nitride (NbN) film preserved single-photon response at visible range and held short reset time ($\tau_{d}\approx5$ ns), as well as large critical currents ($I_C\approx0.5$ mA). This proof-of-principle devices showed great promise for superconducting single-photon detection technology, but demonstrated the presence of a drastic drop of DE in the near-infrared range for wide detectors. Subsequent works addressed the DE of superconducting microstrip single-photon detectors (SMSPDs) by implementing WSi and MoSi superconducting ultra-thin films exhibiting transition temperatures ($T_c$) below 5 K\cite{charaev2020large,chiles2020superconducting}. However, the main downside of low-$T_c$ materials is a demand of sophisticated cryogenic apparatus required for their operation. As a result, such cryogenic systems partly exclude the possibility of wide spreading of the devices. 

 \begin{figure}[t!]
    \centering
    \includegraphics[width=\linewidth]{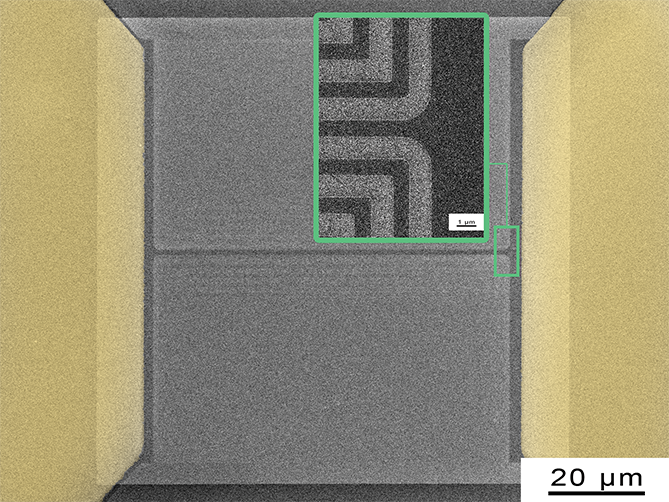}
    \caption{Scanning electron microscope image of device design. Single 75 $\mu$m strip is located in the center between two Ti/Au contact pads (in false color). Inset demonstrates a transition to bridge width with smooth rounding preventing current crowding.}
    \label{structure}
\end{figure}

\begin{figure*}[t!]
    \centering
    \includegraphics[width=\linewidth]{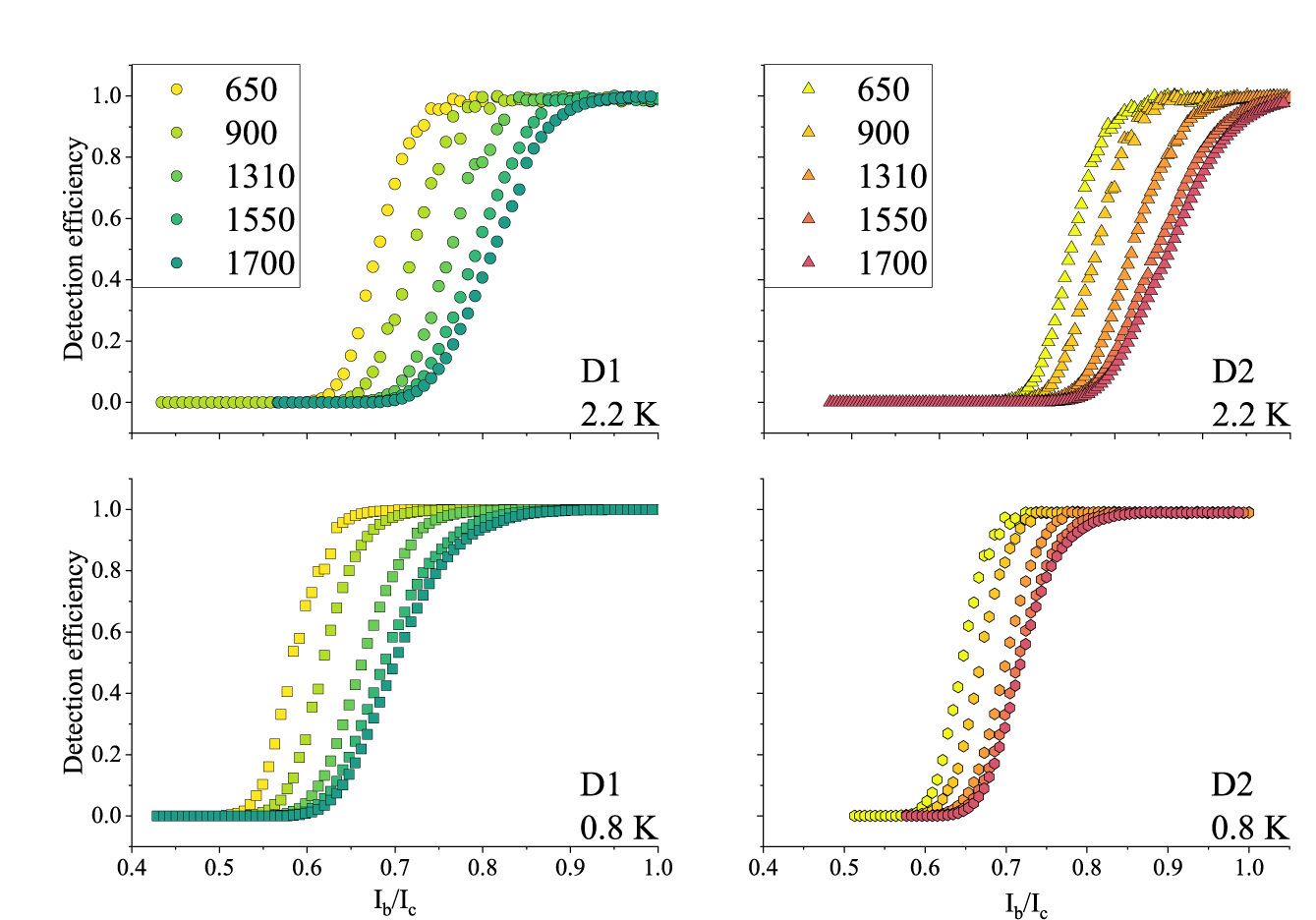}
    \caption{Detection probabilities for two samples with 0.5 and 1 $\mu$m strip width operating at 2.2 and 0.8 K. Curves presented for five wavelengths.}
    \label{dp}
\end{figure*}

\begin{table*}[t!]
\caption{\label{table}Parameters of studied devices. Table contains strip widths measured with scanning electron microscopy. Stated $R_s$ and \textit{RRR} values are measured on D1 and are also employed for D2. $R_s$ is found as maximum value on $R_s(T)$ curve.}
\begin{tabular}{ccccccccc}
\hline
Device  &   Strip width, $\mu$m   & $I_c^{2.2 K}$, $\mu$A   & $I_c^{0.8 K}$, $\mu$A  & $R_s$, k$\Omega$/$\square$ & $T_c$, K  &   \textit{RRR}
& $I_c^{2.2 K}/I_d^{2.2 K}$ & $I_c^{0.8 K}/I_d^{0.8 K}$\\
\hline
\hline
D1      &   0.43 & 30.0    & 35.5 & 1.25 & 8.3 & 0.72 & 0.62 & 0.66\\
D2      &   0.89   & 66.2    & 80.0    & 1.25  &  8.3 & 0.72 & 0.61 & 0.66 \\
\hline
\end{tabular}
\end{table*}

Around the same time it was demonstrated that DE of superconducting single-photon detectors is highly dependent on normal state sheet resistance ($R_s$) of initial films\cite{zolotov2021dependence,semenov2021superconducting}. For increasing sheet resistance saturation plateaus of the dependencies of detection efficiency on bias current extend. This extension, however, remains vulnerable to increase of detector's operation temperature, photon energy, or, as stated above, strip width. While deterioration of saturation in case of a higher operation temperature could be a result of significant drop of current density, and its decline for lower photon energies could be explained by less efficient gap suppression, its change with strip width is not fully understood and requires further experimental investigation\cite{renema2015physics,semenov2021superconducting}. As it is anticipated by Vodolazov photon detection by a microstrip should not depend on its width when one key condition is met. This key condition is the ability of the strip to carry a superconducting current larger than $0.7I_{d}$, where $I_{d}$ is depairing current that is related to critical supervelocity of the Cooper pairs. Therefore, the search of the optimal material for SMSPDs and its operating conditions should reference these key aspects. A good candidate for such role is niobium nitride, which is a conventional material for SSPD fabrication, and is well suited for operation in a wide temperature range thanks to its relatively high critical temperature of the order of 10 K obtained in a thin film. SNSPDs made from niobium nitride films reach system detection efficiencies above 90\% in GM cryocoolers operated at 2.2 K and therefore are convenient both for research purposes, as well as in-field applications\cite{smirnov2018nbn, zolotov2021dependence, esmaeil2021superconducting}. In the field of SMSPDs niobium nitride devices demonstrated proximity to saturation of DE at 1064 nm in 1 and 3 $\mu$m-wide bridges operated at 1.7 K\cite{korneeva2021influence}. At the same time, only near-unity detection efficiency was achieved at 1550 nm for the same conditions. Similar results were independently obtained by implementing helium irradiation for post-fabrication treatment of the detectors and by cooling them to lower temperature of 0.84 K\cite{xu2021superconducting}.  Missing plateaus at 1550 nm in results by Korneeva \textit{et al.} could be attributed to low current density in fabricated structures together with low values of $R_s = 764 \Omega$/$\square$  ($T_c=9$ K) of initial film. Similar performance in work by Xu and co-authors may be explained by low critical temperature ($T_c=6.4$ K at $R_s=1036 \Omega$/$\square$), which might require lower operation temperature than was achieved in the experiment. Considering the above facts our main goal was to experimentally evaluate the performance of SMSPDs made out from niobium nitride film with $R_s\approx1$ k$\Omega$/$\square$ and $T_c\geq8$ K, which are believed to be the key requirements for high-performance SMSPDs. These values are in a optimal correspondence with experimental data for NbN films presented in our previous work\cite{zolotov2021dependence}. As a result, we demonstrate microstrip devices operated in a Gifford -- McMahon (GM) cryocooler and reaching unity DE at telecom wavelengths and beyond. 

Deposition runs of niobium nitride ultra-thin films was performed by reactive magnetron sputtering of niobium target onto sapphire substrates with additional Ti/Au/Si$_3$N$_4$ optical cavity that is often applied for fabrication of practical SNSPDs. During the process substrate was maintained at 300 \textdegree{C}. Before each film deposition the target was presputtered with a closed magnetron shutter for 3 minutes. Over the period of film growth, a negative 250 V RF bias was delivered to the substrate. Argon to nitrogen flow ratio was fixed at 40:15 cm$^3$/min resulting in operating pressure of 4 mTorr. Deposition rate was calibrated using quartz crystal microbalance and controlled via film deposition time. Among deposited samples we chose the film which exhibited the highest critical temperature while holding close to desirable resistance per square. Estimated film thickness was 5 nm. It was found that substrate biasing allowed to increase $T_c$ of deposited film by 0.5 K compared to bias-free deposition which agrees with results by Dane \textit{et al.}\cite{dane2017bias}.  Design of our samples constituted of single bridges with dose-stabilization structures -- Fig \ref{structure}. Devices had the same bridge length of 75 $\mu$m and width of either 0.5 or 1 $\mu$m. Resulting bridge widths together with film parameters are combined in Table \ref{table}. Pattern files were generated using PHIDL package with some additional code supplement for obtaining dose-stabilization structures\cite{mccaughan2021phidl}. Films were patterned using a 30 kV electron-beam lithography with PMMA A3 resist followed by plasma-etching in SF$_6$ with Ar admixture. Fabricated structures were supplemented with Ti/Au contact pads and separated into chips.

\begin{figure}[t!]
    \centering
    \includegraphics[width=\linewidth]{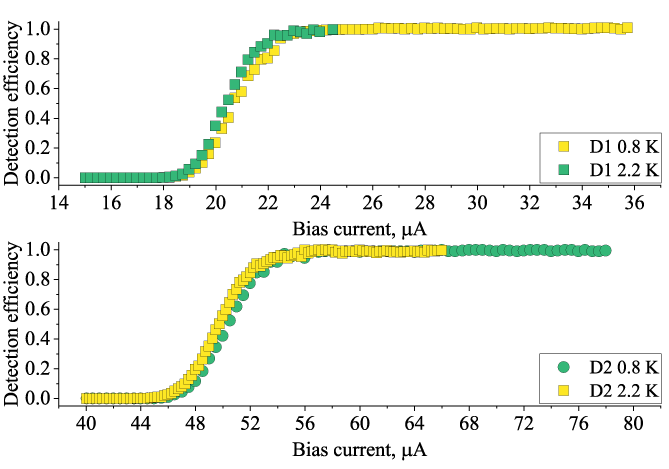}
    \caption{Alteration of DE curves at 650 nm for D1 (top) and D2 (bottom). At lower temperatures the curves extend to higher bias current values without any significant shift in terms of curve center values.}
    \label{650}
\end{figure}

In order to evaluate the change of DE not only for two studied bridge widths, but also for two operation temperatures, measurements of fabricated devices were performed at 2.2 and 0.8 K in a GM closed-cycle refrigeration cryostat with additional sorption cooler module.  Devices were installed on PCBs using varnish and wire bonding. Operating temperature of the detectors was measured with a diode thermometer installed in the same setup. Signal from devices was led from the cryostat via coaxial cables to the room temperature bias-tee followed by amplifier chain. Due to low inductance of the devices that leaded to latching a 470 nH SMD inductor coil was connected in series with each detector. Measurement routine consisted of recording photocounts dependence on bias current at various wavelengths (650, 900, 1310, 1550, 1700 nm) fed into the cryostat via optical fiber. Devices were flood-illuminated while output power of the laser was attenuated to establish similar counting rate for each device at each wavelength. Radiation was generated by supercontinuum laser source (repetition rate was 20 MHz) with acousto-optic tunable filter providing a typical wavelength peak width of less than 5 nm. To avoid high heat loads of the cryogenic stage operating temperature was continuously monitored. Figure \ref{dp} presents performance of two selected from fabricated batch devices with different strip width values. Both devices reach unity DE at operating temperature of 2.2 K in a wide spectral range. When detectors are cooled down to 0.8 K they both demonstrate high saturation level of DE with saturation plateaus beginning at approximately $0.75I_c$ -- $0.85I_c$ depending on the wavelength and width. To better visualize how DE evolves for various conditions we separately plot obtained curves of each detector at incident wavelength of 650 nm at both operating temperatures --- Figure \ref{650}. As it follows from the graphs, when plotted in absolute values of the bias current, dependencies taken at 0.8 K are a direct continuation of those obtained at 2.2 K.

To find out how obtained results relate to Vodolazov theory we estimate the relation between operating currents of our detectors and depairing values. For that an additional experiment was made to measure critical temperature of fabricated devices along with residual resistivity ratio \textit{RRR} found as $ R_s^{300 K}/R_s^{20 K}$. For D1 we find $I_c/I_{dep}$ equal to 53.5 and 48.6 $\mu$A at 0.8 and 2.2 K correspondingly. And for D2 --- 120.3 $\mu$A at 0.8 K and 109.4 $\mu$A at 2.2 K. The calculations were made under an assumption that diffusion coefficient in our films is equal to 0.5 cm/s$^2$. It should be highlighted that obtained critical temperature of fabricated devices was within 0.1 K lower than that of as-deposited film. Interestingly, obtained \textit{RRR} value was approximately 10\% higher than the typical value for our films with similar $R_s$ but deposited without RF bias. For convenience, measured values are added to Table \ref{table}.

It appears from what has been demonstrated that NbN films preserve the desired properties for fabrication of superconducting microstrip single-photon detectors that demonstrate saturated detection efficiency at telecom wavelengths and beyond. Such achievement is attributed to implementation of a film with high sheet resistance value that simultaneously preserve high critical temperature, which also allows device operation in conventional cryocoolers. Described technology may streamline fabrication process of SMSPDs with the help of photolithography and still offer high performance. It is also believed that SMSPDs could become a helpful testbed for evaluation of the optimal thin superconducting films' parameters. 
\bibliographystyle{unsrtnat}
\bibliography{synapse}
\end{document}